\documentclass[prd,eqsecnum,twocolumn,amsfonts,showpacs]{revtex4}
\newcommand{\beq}{\begin{equation}}
\newcommand{\eeq}{\end{equation}}
\newcommand{\beqs}{\begin{eqnarray}}
\newcommand{\eeqs}{\end{eqnarray}}

\newcommand{\gsim}{\mathrel{\raisebox{-
.6ex}{$\stackrel{\textstyle>}{\sim}$}}}

\newcommand{\Dslash}{D\hspace{-0.09in}\slash}

\begin{document}

\title{On the $\pi$ and $K$ as $q \bar q$ Bound States and Approximate
Nambu-Goldstone Bosons}

\author{Shmuel Nussinov$^{a,b}$}

\author{Robert Shrock$^c$}

\affiliation{(a)
School of Physics and Astronomy, Tel Aviv University, Tel Aviv, Israel}

\affiliation{(b)
Schmid College of Science, Chapman University, Orange, CA 92866} 

\affiliation{(c)
C.N. Yang Institute for Theoretical Physics, Stony Brook University,
Stony Brook, NY 11794}

\begin{abstract}

We reconsider the two different facets of $\pi$ and $K$ mesons as $q \bar q$
bound states and approximate Nambu-Goldstone bosons.  We address several
topics, including masses, mass splittings between $\pi$ and $\rho$ and between
$K$ and $K^*$, meson wavefunctions, charge radii, and the $K-\pi$ wavefunction
overlap.

\end{abstract}

\pacs{11.30.Rd, 12.38.-t, 12.39.Jh, 12.40.Yx}

\maketitle

\section{Introduction}

Quantum chromodynamics (QCD) is most predictive in the perturbative,
short-distance regime. Yet our understanding of long distance, non-perturbative
properties of QCD keeps improving. Lattice QCD has been used to compute the
hadronic spectrum and matrix elements for weak transitions, the $N_c \to
\infty$ limit and new variants thereof have been widely applied, and
heavy-quark symmetry has helped to elucidate heavy-heavy $Q\bar Q$ and
heavy-light $Q\bar q$ hadrons. The ${\rm SU}(2)_L \times {\rm SU}(2)_R$ global
chiral symmetry and the spontaneous breaking of this symmetry to the vectorial,
isospin SU(2)$_V$, which are mandatory in QCD \cite{casher}-\cite{csblgt},
underlie isospin symmetry, the fact that the pions are light almost
Nambu-Goldstone bosons (NGB's) \cite{njl,goldstone}, and the usefulness of
chiral perturbation theory. Another specific advance was the resolution of the
U(1)$_A$ problem by instantons \cite{thooft76}, thus explaining why the $\eta'$
is not light.

The remarkable success of the nonrelativistic (``naive'') quark model (NQM)
treating the $u$, $d$, and $s$ quarks as nonrelativistic, spin-1/2 fermions is
of great interest \cite{nqm}-\cite{gm}. This model dictated the low-lying
flavor SU(3) multiplets and many aspects of their electroweak interactions.
Spontaneous chiral symmetry breaking (S$\chi$SB) generates dynamical,
constituent masses of order the QCD scale $\Lambda_{QCD} \simeq 250$ MeV for
the light, almost massless $u$ and $d$ quarks and increments by this amount
the hard Lagrangian mass of the $s$ quark, $m_s \simeq 120$ MeV to produce a
constituent $s$ quark mass.  Our results are not sensitively dependent upon the
values of the constituent quark masses; we use the values \cite{pdg,qmass}
\beq
M_u = M_d \equiv M_{ud} \simeq 340  \ \ {\rm MeV}
\label{mud}
\eeq
and
\beq
M_s \simeq 470 \ \ {\rm MeV} \ . 
\label{ms}
\eeq
The dynamical mass generation of the constituent quark masses in QCD can, for
example, be shown via analysis of the Dyson-Schwinger equation for the quark
\cite{sdeq}.  One may roughly characterize the range of the QCD interactions
responsible for the $\langle \bar q q\rangle$ condensate as $r_0$.  The
effective size of a constituent quark, consisting of the bare valence quark and
its entourage of gluons and $q \bar q$ pairs, is expected to be of order
$r_0$. A small $r_0$, less than 0.2 fm, say, then yields constituent quarks of
that size moving inside a hadron of size approximately 1 fm under the influence
a smooth confining potential, making the NQM plausibly justified.

Some of the mechanisms suggested for generating spontaneous chiral symmetry
breaking - in particular, Nambu-Jona Lasino-type (NJL) models
\cite{njl,njlafterqcd}, and those involving instantons
\cite{thooft76,caldicdg,instanton2} - can have relatively short range. However
in the approach of Casher \cite{casher} and Banks and Casher \cite{bc},
S$\chi$SB results from confinement. In this approach there is no separation of
scales between the constituent quark size and hadron sizes.

The almost massless Nambu-Goldstone pion - the other consequence of spontaneous
chiral symmetry breaking - generates arguably the single most serious
difficulty for the NQM, namely what has been called the ``$\rho$--$\pi$
puzzle'' (e.g., \cite{pagels}) This is the challenge of simultaneously
explaining the pion as a $q \bar q$ bound state and an approximate NGB, and
relating it to the $\rho$.  There is an analogous, although less severe,
problem for the $K$ and $K^*$. Although this difficulty is most clearly
manifest within the NQM, it transcends this nonrelativistic model.  Thus, also
the original MIT bag model with relativistic quarks confined in a spherical
cavity requires large hyperfine interactions to try to split the masses of the
$\pi$ and $\rho$ and, like the NQM, fails to explain the almost massless pions
\cite{mitbag}.  To get a sufficiently light pion in the MIT bag model, it is
necessary to argue that subtantial contributions due to the fluctuations in the
center-of-mass should be subtracted \cite{dj} (see also
\cite{goldhay,chiralbag}).

In this paper we shall revisit the problem of understanding the dual nature of
the pion (and kaon) as $q \bar q$ bound states and as collective, almost
massless Nambu-Goldstone bosons.  The organization of the paper is as follows.
In Section II we elaborate on the $\rho$--$\pi$ puzzle in the context of the
nonrelativistic quark model.  In Section III we present a heuristic picture
that gives some new insight into this puzzle by helping to explain the $\pi$ as
both a $q \bar q$ bound state and an approximate Nambu-Goldstone boson.  In
Section IV we consider the $K-\pi$ transition form factor, $f_+(q^2)$ .  Its
deviation from unity at vanishing momentum transfer is governed by the
Ademollo-Gatto theorem \cite{ag}, and analogous deviations from heavy quark
universality are derived in a general context . In Section V we comment on the
systematics of quark mass differences inferred from $(Q \bar s)$ and $(Q \bar
q)$ mesons, where $q=u$ or $d$. This has been elaborated independently by
Karliner and Lipkin \cite{kl}. Still, we feel that it is of sufficient interst
to present it here from our point of view.

\section{The $\rho$-$\pi$ Puzzle}

   Several pieces of data suggest that the pion, which is lighter than the
$\rho$ by approximately $640$ MeV, is otherwise rather similar to the $\rho$,
as expected in the nonrelativistic quark model for the ${}^1S_0$ pseudoscalar
partner of the ${}^3S_1$ vector meson.  These data can be summarized as
follows:

\begin{itemize}

\item

One measure of the size of a hadron is provided by the magnitude of the charge
radius. The charge radii of the $\pi^+$ and $K^+$ are given by \cite{pdg}
\beq
(\langle r^2 \rangle_{\pi^+})^{1/2} = 0.672 \pm 0.008 \quad {\rm fm}
\label{rsqpip}
\eeq
and
\beq
(\langle r^2 \rangle_{K^+})^{1/2} = 0.560 \pm 0.031 \quad {\rm fm} \ . 
\label{rsqkp}
\eeq
These are rather similar and, indeed, are not very different from the
charge radius of the proton,
\beq
(\langle r^2 \rangle_p)^{1/2} = 0.875 \pm 0.007 \quad {\rm fm} \ . 
\label{rsqp}
\eeq

\item

Similar $\pi$ and $\rho$ sizes and a somewhat smaller kaon are suggested by the
total cross sections on protons at a typical laboratory energy above the
resonance region.  Averaged over meson charges, at a lab energy $E_{lab}=10$
GeV, these are \cite{pdg}:
\beq
\sigma_{\pi N} \simeq 26.5 \quad {\rm  mb}
\label{sigma_pin}
\eeq
and
\beq
\sigma_{K N} \simeq 21 \quad {\rm mb} \ . 
\label{sigma_kn}
\eeq
Although one obviously does not have beams of $\rho$ mesons available
experimentally, owing to the very short lifetime of the $\rho$, it is possible
to estimate what the cross section for $\rho-N$ scattering would be if one did
have such beams.  Diffractive $\rho$ production data and vector meson dominance
yield the estimate \cite{bauerpipkin}
\beq
   \sigma_{\rho N} \simeq  27 \pm 2 \ \ {\rm mb} \ . 
\label{sigma_rhon}
\eeq
This cross section is essentially the same, to within the experimental and
theoretical uncertainties, as $\sigma_{\pi N}$ at the same energy (and these
are approximately equal to $(2/3)\sigma_{NN}$ at this energy).

\item

The nonrelativistic quark model was able to fit the measured values of the
proton and neutron magnetic moments $\mu_p = 2.793 \mu_N$ and $\mu_n = -1.913
\mu_N$, (where $\mu_N = e/(2m_p)$) and the ratio $\mu_p/\mu_n \simeq -3/2$, as
well as the values of the hyperon magnetic moments, in terms of Dirac magnetic
moments of constituent quarks. It also explained decays such as $\omega \to
\pi^0 + \gamma$ and $\rho \to \pi \gamma$ as quark spin flip ${}^3S_1 \to
{}^1S_0$ electromagnetic transitions.  The optimal overlap of the $\rho$ and
$\pi$ wavefunctions implied by this confirms the similarity of the vector and
pseudoscalar meson ground state wavefunctions.

\item

The amplitudes for semileptonic $K_{\ell 3}$ decays involve the vector part of
the weak $|\Delta S|=1$ current and contain the product of $V_{us}$ with the
$f_+(q^2)$ transition form factor. In the limit of SU(3) flavor symmetry
$m_u=m_d=m_s$, so that $m_K=m_{\pi}$, the conserved vector current (CVC)
property implies that $f_+(0)=1$. Experimentally, $f_+(q^2=0)$ is very close to
unity.  The success of these fits implies almost optimal overlap between the
wavefunctions of the pion and kaon.

\end{itemize}

In the nonrelativistic quark model, one can rewrite the two-body
quark-antiquark Hamiltonian as an effective one-body problem with the usual
reduced mass
\beq
\mu_{i \bar j} = \frac{M_iM_j}{M_i+M_j}
\label{mu}
\eeq
for the $q_i \bar q_j$ pseudoscalar meson. (The context will make clear where
the notation $\mu$ refers to a magnetic moment and where it refers to a reduced
mass.)  The corresponding bound-state wave function is denoted $\psi_\pi({\bf
r})$, $\psi_K({\bf r})$, etc., where ${\bf r} =
{\bf r}_{q_i}-{\bf r}_{\bar q_j}$ is the relative coordinate in the bound
state.  With the above-mentioned typical values $M_{ud}=340$ MeV and $M_s =
470$ MeV, one has
\beqs
\mu_\pi =  \frac{M_{ud}}{2} = 170 \ \ {\rm MeV}
\label{mupi}
\eeqs
and
\beq
\mu_K = \frac{M_{ud} M_s}{M_{ud}+M_s} = 200 \ \ {\rm MeV} \ , 
\label{muk}
\eeq
where it is understood that the choices for the input values of the constituent
quark masses in these formulas depend somewhat on the method that one uses to
infer their values \cite{qmass}. In the nonrelativistic quark model, since a
bound state involving a larger effective reduced mass is expected to be
smaller, one has some understanding of the fact that $\sqrt{\langle r^2
\rangle_{K^+}} \simeq 0.83 \sqrt{\langle r^2 \rangle_{\pi^+}}$. The deviation
of $f_+(0)$ from unity is also in accord with this difference of reduced
masses.

For heavy $Q \bar Q$ quarkonium systems one can use the nonrelativistic
Schr\"odinger equation to describe a number of properties of the bound states
\cite{qr,gm}.  This use is justified by the fact that in the $c \bar
c$ and $b \bar b$ systems the respective heavy quark masses $m_c \simeq 1.3$
GeV and $m_b \simeq 4.3$ GeV are large compared with $\Lambda_{QCD}$, and the
asymptotic freedom of QCD means that $\alpha_s$ gets small for such mass
scales.  The hyperfine splitting in these $\bar Q Q$ systems, being
proportional to $\alpha_s/m_Q$, is small.

For light $\bar q q$ systems, however, the situation is different.  Let
us denote
\beq
\langle 0 | J^j_\lambda |\pi^k(p)\rangle = i f_\pi \delta^{jk} p_\lambda \ ,
\label{matrixelement}
\eeq
where $j,k$ are isospin indices and $J_\lambda$ is the weak charged current, so
that $\langle 0 | J^{1-i2}_\lambda | \pi^+(p) \rangle = i f_\pi p_\lambda$.
Similarly, $\langle 0 | J^{4-i5}_\lambda | K^+(p) \rangle = i f_K
p_\lambda$. Experimentally \cite{pdg},
\beq
f_\pi = 92.4 \ \ {\rm MeV}, \quad\quad f_K = 113 \ \  {\rm MeV} \ . 
\label{fpik}
\eeq
Analogous constants enter in the leptonic decays of
the vector mesons. The rate for the decay $M^+_{i \bar j} \to
\ell^+ \nu_\ell$, where $\ell=\mu$ or $e$, is
\beqs
& & \Gamma(M_{i \bar j} ^+ \to \ell^+ \nu_\ell) =
\frac{|V_{ij}|^2 G_F^2 f_{M_{i \bar j}}^2 m_{M_{i \bar j}}m_\ell^2 }{4 \pi}
\, \left [ 1 - \frac{m_\ell^2}{m_{M_{i \bar j}}^2} \right ]^2 \cr\cr
& & 
\label{pienu}
\eeqs
where here $V_{ij}=V_{ud}$ for $M_{u \bar d}^+=\pi^+$ and $V_{us}$ for
$M_{u \bar s}^+=K^+$. Since $M_{i \bar j}$ is a $q_i \bar q_j$ bound state,
this rate is proportional to $|\psi(0)|^2$.  With the normalization of
$\psi(r)$ in the nonrelativistic quark model determined by the condition $\int
d^3 r |\psi(r)|^2 = 1$, it follows that for a given ${}^1S_0$ or ${}^3S_1$ $q_i
\bar q_j$ meson $M$,
\beq
    f_M \propto \frac{|\psi_M(0)|}{m_M^{1/2}} \ . 
\label{vrw}
\eeq
Hence,
\beq
\frac{|\psi_K(0)|}{|\psi_\pi(0)|} = \frac{f_K}{f_\pi} \left ( \frac{m_K}{m_\pi}
\right )^{1/2} = 2.3 \ . 
\label{psiratio}
\eeq
The difficulty of deriving this ratio from the NQM was noted early on as the
van Royen-Weisskopf ``paradox'' \cite{vrw}.

      Conventionally, in the context of the quark model, the $\rho$--$\pi$ mass
difference was explained by means of a very strong chromomagnetic, i.e., color
hyperfine (chf) splitting between these particles.  The similar, although
smaller, mass difference between the $K^*$ and $K$ was also explained by this
color hyperfine interaction. Taking account of color, the Hamiltonian for the
color hyperfine (chromomagnetic) interaction has the form
\beq
H_{chf} =  \frac{v_{chf}(r)}{M_i M_j}( {\vec \lambda}_i \cdot {\vec \lambda}_j)
({\vec \sigma}_i \cdot {\vec \sigma}_j) \ , 
\label{hcf}
\eeq
where the function $v_{chf}(r)$ involves the overlap of the interacting
constituent (anti)quarks.  Recall that the product ${\vec \sigma}_i \cdot {\vec
\sigma}_j$ is equal to 1 and $-3$ times the identity matrix ${\mathbb I}_{2
\times 2}$ when the $q_i$ and $\bar q_j$ spins are coupled to $S=1$ and $S=0$,
respectively.  Similarly, the product of SU(3)$_c$ color matrices $ {\vec
\lambda}_i \cdot {\vec \lambda}_j$ is equal to $-16/3$ and $-8/3$ times
${\mathbb I}_{3 \times 3}$ if the colors are coupled as $3 \times \bar 3 \to 1$
(meson) and $3 \times 3 \to \bar 3$ (baryon), respectively. The resultant
$1:(-3)$ ratio of mass shifts in $S=1$ and $S=0$ $q \bar q$ mesons or quark
pairs in baryons and the $1:(1/2)$ ratio of the color factor for $q \bar q$
mesons versus $qq$ interactions in baryons yield good fits to meson and baryon
masses.  The dependence of $H_{chf}$ on $1/(M_iM_j)$ is also important for this
successful fit.  In the NQM as applied here, $v_{chf}(r) \propto \delta^3({\bf
r})$, so that the color hyperfine shifts evaluated to first order in $H_{chf}$
are proportional to $|\psi(0)|^2$ (where the subscript $M$ on $\psi$ is
suppressed in the notation) \cite{nqm}.  This is analogous to the hyperfine
splitting in hydrogen, which is also proportional to the square $|\psi(0)|^2$
of the hydrogenic wavefunction at the origin.  We focus here on the ${}^1S_0$
and ${}^3S_1$ isovector mesons, i.e., the $\pi$ and $\rho$, absorb the color
factor into the prefactor and thus write, for the energy due to $H_{chf}$,
\beq
E_{hcf} = \frac{A \langle {\vec \sigma_i} \cdot {\vec \sigma_j}\rangle
|\psi(0)|^2}{M_i M_j} \ . 
\label{ehcf}
\eeq
The meson mass to zeroth order in $H_{chf}$ is denoted $m_0$.  Then
the physical masses are
\beq
m_\rho = m_0+\frac{A|\psi(0)|^2}{M_{ud}^2}
\label{mrhoshift}
\eeq
and
\beq
m_\pi=m_0-\frac{3A|\psi(0)|^2}{M_{ud}^2} \ . 
\label{mpishift}
\eeq
Equivalently,
\beq
m_0 = \frac{3m_\rho + m_\pi}{4}
\label{mvals}
\eeq
and
\beq
A = \frac{(m_\rho - m_\pi)M_{ud}^2}{4|\psi(0)|^2} \ . 
\label{arel}
\eeq
Numerically, $m_0=620$ MeV and the color hyperfine splitting is
\beq
(\Delta E)_{chf} = m_\rho - m_\pi = \frac{4A|\psi(0)|^2}{M_{ud}^2} = 640
\ \ {\rm MeV} \ . 
\label{deltae}
\eeq

The color hyperfine interaction also plays an important role in splitting the
$K$ and $\pi$, since their mass difference, $m_K - m_\pi \simeq 360$ MeV,
exceeds, by about a factor of 3, the difference in current-quark masses, $m_s -
m_d \simeq 120$ MeV. In the NQM, this is attributed to the fact that the color
hyperfine interaction energy $-a/M_{ud}^2$ for the $\pi$ (with $a > 0$) is
negative and larger in magnitude than the corresponding energy $-a/(M_{ud}M_s)$
for the $K$, since $M_s > M_{ud}$.

The large size of the splitting (\ref{deltae}) shows that the color hyperfine
interaction is not a small perturbation on the zeroth-order Hamiltonian value,
$m_0$. Furthermore, the strongly attractive, short-range color hyperfine
interaction has the effect of contracting the pion to a size substantially
smaller than the size indicated by experimental data on the charge radius and
$\sigma_{\pi N}$ scattering cross section.  Moreover, since $v_{chf}(r) \propto
\delta^3({\bf r})$, which is clearly sensitive to short-distance interactions
between quarks, there is a problem of internal consistency when one uses a
color hyperfine interaction in the context of the nonrelativistic quark model,
since at short distances, because of asymptotic freedom, the light quarks
behave in a relativistic quasi-free manner with their small, current-quark
masses, not as nonrelativistic, massive, constituent quarks.

The fact that $M_{ud}/M_s \simeq 0.7$ has two countervailing effects on the
relative charge radii of the $\pi^+$ and $K^+$.  First, since the reduced mass
$\mu_K$ is slightly greater than $\mu_\pi$ (c.f. Eqs. (\ref{mupi}) and
(\ref{muk})), it is plausible that the corresponding meson could be somewhat
smaller. Yet the very important attractive color hyperfine interaction should 
make $\sqrt{\langle r^2 \rangle_{K^+}}$ larger than 
$\sqrt{\langle r^2 \rangle_{\pi^+}}$.

We proceed to discuss some properties of the color hyperfine interaction
further.  Since an attractive interaction involving a $\delta^3({\bf r})$
function potential is inconsistent in the nonrelativistic constituent quark
model, we model $v_{chf}(r)$ as a spherical square well of depth $V_0$ and
radius $r_0$.  The dimensionless quantity fixing the number of bound states is
the ratio of the strength $V_0$ of a potential to the kinetic energy of a
particle bound by this potential in a region of size $r_0$, namely $E_{kin}
\simeq p^2/(2\mu) = \pi^2/(2\mu r_0^2)$ ($E_{kin} = p = \pi/r_0$,
relativistically). The ratio $V_0/E_{kin} = 2\mu V_0 r_0^2/\pi^2$. Since $\mu$
is determined by Eq. (\ref{mu}), we thus fix the product $V_0r_0^r$.  One knows
from general QCD theory that at distances $r << 1/\Lambda_{QCD}$, the static
quark potential has the Coulombic form
\beq
V_{q \bar q} = \frac{C_{2f}\alpha_s}{r} \quad  {\rm for} \ \
r << \frac{1}{\Lambda_{QCD}} \ , 
\label{vcoulombic}
\eeq
where $C_{2f}=4/3$ is the quadratic Casimir invariant for the fundamental
representation of SU(3)$_c$ and the logarithmic dependence of the running
$\alpha_s$ on $r$ is left implicit.  At distances of order $1/\Lambda_{QCD}$,
$V_{q \bar q}$ has a linear form resulting from the chromoelectric flux tube
joining the $q$ and $\bar q$,
\beq
V_{q \bar q} = \sigma r \quad {\rm for} \ \ r \sim \frac{1}{\Lambda_{QCD}} \ , 
\label{vlinear}
\eeq
where $\sigma = 1/(2 \pi \alpha') \simeq (400 \ {\rm MeV})^2$ is the string
tension with $\alpha'$ the Regge slope.  To illustrate the nature of the
$\rho$-$\pi$ puzzle, let us consider an infinite square well potential, which
provides a simple model of confinement. The (unit-normalized) ground state
wavefunction is
\beq
\psi(r) = \left (\frac{\pi}{2r_0^3} \right )^{1/2} \left ( \frac{\sin pr}{pr}
\right ) 
\label{psisquarewell}
\eeq
where
\beq
p=\frac{\pi}{r_0} \ . 
\label{kvalue}
\eeq
With this potential, one has
\beqs
\langle r^2 \rangle & = & \int \, r^2 |\psi(r)|^2 d^3r \cr\cr
     & = & \left ( \frac{1}{3}-\frac{1}{2\pi^2} \right ) r_0^2 \ , 
\label{rsq_squarewell}
\eeqs
so that $\sqrt{\langle r^2 \rangle} = 0.532 r_0$. The measured value of
$\langle r^2 \rangle_\pi$ then determines $r_0 = 1.26$ fm.  Denoting
$\langle r^2 \rangle \equiv d^2$, one can write
\beq
|\psi(0)|^2 = \frac{c}{d^3}
\label{psisqd}
\eeq
with $c$ a constant.  In this model,
\beq
c = \frac{\pi}{2}\left ( \frac{1}{3} - \frac{1}{2\pi^2} \right )^{3/2} = 0.236
\label{cvalue}
\eeq
Substituting the value of $|\psi(0)|^2$ from Eq. (\ref{psisquarewell}) into
Eq. (\ref{deltae}), we find
\beq
(\Delta E)_{chf} = \frac{2\pi A}{M_{ud}^2 r_0^3}
\label{deltaesquarewell}
\eeq
so that
\beq
A = \frac{(m_\rho - m_\pi)M_{ud}^2 r_0^3}{2\pi} = 3.1 
\label{avalue}
\eeq
Thus, both the large shift in Eq. (\ref{deltae}) and the rather large value of
the coefficient $A$ show that the NQM treatment of the very strong, short-range
hyperfine interaction as a perturbation is not really justified.  As could be
expected on general grounds, such a strong short-range interaction has the
effect of producing a pion wavefunction that is smaller in spatial extent than
is experimentally observed.  In effect, the pion - which, by definition, is the
ground state in the attractive ${}^1S_0$ pseudoscalar channel - is
``swallowed'', i.e., squeezed into a contracted state of radius much smaller
than that of the $\rho$.  The wavefunction for the $\rho$ itself slightly
expands relative the original common unperturbed $\pi$ and $\rho$
wavefunctions, due to the repulsive color hyperfine interaction in the
${}^3S_1$ vector channel (which is 1/3 as strong as the attraction in the
${}^1S_0$ channel).  The NQM puzzle of a very light pion thus extends also to
its expected much smaller size.  In general, any extra, attractive, potential
that binds the pion more strongly than the $\rho$ yields a $\pi$ that is
smaller than the $\rho$.
The only way to maintain a common shape for the $\rho$ and $\pi$ wavefunctions
in a nonrelativistic potential model framework is to have $v_{chf}({\bf r})$
constant as a function of $r$, which is very different from the NQM's form
$v_{chf} \propto \delta^3({\bf r})$.

\section{A Physical Picture of Approximate Nambu-Goldstone Bosons}

NJL-type models do succeed in producing a massless or nearly massless pion in a
bound-state picture, as was shown first via a solution of the Bethe-Salpeter
equation in the ${}^1S_0$ channel in the original work by Nambu and Jona
Lasinio \cite{njl} (with an appropriate reinterpretation of the four-fermion
operator as involving quarks rather than nucleons in a modern context
\cite{njlafterqcd}).  However, the coupling of the four-fermion operator is not
calculated directly from the underlying QCD theory.  Furthermore, this
four-fermion operator posits that the spontaneous chiral symmetry breaking is a
contact interaction, and thus does not directly include the physically
appealing mechanism for S$\chi$SB as being a consequence of helicity reversal
due to confinement \cite{casher}.

An important insight for understanding the pion as a $q \bar q$ bound state and
also an approximate Nambu-Goldstone boson has been the argument by Brodsky and
Lepage that the physical pion state contains not just the valence $|q \bar
q\rangle$ state, but large contributions from higher Fock states such as $|q
\bar q + ng \rangle$, $|q \bar q q \bar q \rangle$, $|q \bar q q \bar q + ng
\rangle$ ($g=$ gluon), etc. \cite{bl}. These higher Fock space states can
account for much of the size of the physical pion.  This view is in accord with
the Goldstone phenomenon in condensed matter physics, where a Goldstone
excitation is a collective state (e.g., a quantized spin wave or magnon in the
case of a ferromagnet).  This insight has been deepened with further work using
the light front formalism \cite{lfrev}.  Recently, Brodsky and de T\'eramond
have used AdS/QCD methods to calculate hadron masses including $m_\pi$, and
also $f_\pi$, $\langle r^2 \rangle_{\pi^+}$, and the pion electromagnetic form
factor $F_{\pi^+}(q^2)$ in the spacelike and timelike regions \cite{bt}.  

Another approach is provided by approximate solutions to Dyson-Schwinger and
Bethe-Salpeter equations.  In addition to phenomenological four-fermion
NJL-type kernels for the Bethe-Salpeter equation \cite{njl,njlafterqcd}, these
have involved quark-gluon interactions in an effort to model QCD
\cite{sdeq,bs}.  These equations capture some of the relevant physics, although
they do not directly include effects of confinement or nonperturbative effects
due to instantons.  Confinement means that both quarks and gluons have maximum
wavelengths, i.e., minimum bound-state momenta, which affect chiral symmetry
breaking \cite{lmax}.  (Solutions of Dyson-Schwinger and Bethe-Salpeter
equations have also been used to investigate the dependence of the hadron mass
spectrum on the number of light flavors in a general asymptotically free,
vectorial nonabelian gauge theory \cite{nfcr}.)

Thus, there has been continual progress in understanding the pion (and kaon) as
both a $q \bar q$ bound state and an approximate Nambu-Goldstone boson. Here we
would like to present a rather simple heuristic picture of this physics which,
we believe, contributes further to this progress. For technical simplicity, we
restrict ourselves to the large-$N_c$ limit, in which quark loops have a
negligibly small effect.  In this limit a simple proof that spontaneous chiral
symmetry breaking occurs was constructed by showing that the 't Hooft anomaly
matching conditions \cite{thooftanom} for massless $u$ and $d$ quarks must be
realized in the physical spectrum via a massles Nambu-Goldstone pion rather
than massless nucleons \cite{cw}. To motivate our picture, we note several
elements that were missing in the nonrelativistic quark model approach to the
$\rho$-$\pi$ puzzle:

\bigskip
\bigskip

(i) We need a natural mechanism for producing a sufficiently strong $q \bar q$
interaction in the ${}^1S_0$ channel to reduce the mass of the bound
state so that, up to electroweak corrections, it vanishes in the limit of zero
current-quark masses.

\bigskip
\bigskip

(ii) We would like the same physical picture to explain how the pion is both a
$q \bar q$ bound state and an approximate Nambu-Goldstone boson whose
masslessness follows from the spontaneous breaking of the ${\rm SU}(2)_L \times
{\rm SU}(2)_R$ global chiral symmetry down to the diagonal, vectorial SU(2)$_V$
and whose interactions involve derivative couplings, which vanish as $q_\mu \to
0$.  As part of this, it is desirable that the picture should yield the
Gell-Mann-Oakes-Renner (GMOR) formula for the pion (and kaon) mass
\cite{gmor,dashen}.

\bigskip
\bigskip

(iii) We would like to resolve the Van-Royen Weisskopf ``Paradox'' \cite{vrw}
and explain how the quark wavefunction of the pion at the origin,
$\psi_{\pi}(0)$, can be $\propto {m_{\pi}}^{1/2}$ and thus be consistent with a
finite value of $f_\pi$ in the chiral limit where $m_{\pi} \to 0$.

\bigskip
\bigskip

(iv) Finally, we would like to understand how an approximate Nambu-Goldstone
boson such as the pion, which appears quite different from other hadrons, can
have, as indicated by experiment, roughly the same size as these other hadrons.

\bigskip
\bigskip

We next present our new picture and show how it addresses these questions.  As
is well-known, if a quark has zero current-quark mass, the covariant derivative
$\bar q \Dslash q$ in the QCD Lagrangian, preserves chirality.  A dynamical,
constituent quark mass can be generated via an approximate solution of the
Dyson-Schwinger equation for the quark propagator.  In the one-gluon exchange
approximation one finds a nonzero solution for the effective quark mass $M$ if
$C_{2f}\alpha_s = (4/3)\alpha_s \gsim O(1)$. In this framework, the dynamical
quark mass $M$ is thus the consequence of a sufficiently strong quark-gluon
coupling at the relevant scale, $\alpha_s(\mu)$ at $\mu \sim \Lambda_{QCD}$.
This dynamical quark mass can also be seen to result from the helicity reversal
due to confinement \cite{casher}.  These two approaches can also be seen to
connect with the NJL-type analysis, with $M \simeq G \langle \bar q q \rangle
\simeq \langle \bar q q \rangle/(2 \pi f_\pi^2)$, where $G$ denotes the NJL
four-fermion coupling.  If one restricts oneself to a quenched approximation in
which there are no quark loops, then the presence of higher Fock space states
with $q \bar q$ pairs inside a meson with its valence constituent quarks can be
regarded as being due to a kind of {\it zitterbewegung} motion of these valence
quarks.  Light-quark mesons which are not approximate NGB's, such as the
$\rho$, can be modelled satisfactorily as being composed simply of a
constituent quark and constituent antiquark. The effective size of this
constituent $q \bar q$ bound state is of order $1/M_{ud}$.  The application of
the nonrelativistic constituent quark model to such mesons is reasonable, with
the constituent quarks moving in an approximately nonrelativistic fashion under
the assumed confining potential, with a weakly repulsive hyperfine interaction
for the $\rho$ meson.

   In the pion, however, the interaction at all scales is (strongly)
attractive.  This is manifested in the Euclidean pseudoscalar correlator.  We
recall that several general properties of hadrons have been understood on the
basis of Euclidean correlation function inequalities
\cite{dwineq}-\cite{ineqrev}.  Let us consider the Euclidean correlation
function for a pseudoscalar $q \bar q$ bound state,
\beq P(x,y) = \langle [\bar u(x)\gamma_5 d(x)][\bar d(y)\gamma_5 u(y)] \rangle
\ .
\label{pxy}
\eeq
Performing the Gaussian fermionic (Grassman) integration in the path integral
yields
\beq
P(x,y) = \int d\mu(A_{\nu}(x)) S(A)^\dagger(x,y) S(A)(x,y) \ ,
\label{P}
\eeq
where $d\mu(A_{\nu}(x))$ is the positive measure of the Euclidean path
integral, including the $e^{S_G}$ factor from the gauge part of the action,
where
\beq
S_G = \frac{1}{4}G_{\mu\nu}G^{\mu\nu} \ , 
\label{sg}
\eeq
and  the fermionic determinant is absent in the quenched approximation
used here.  In Eq. (\ref{P}), $S(A)(x,y)$ is the propagator of the quark (a
light $u$ or $d$ quark) moving from the initial position $y$ to the final
position $x$, in the presence of the background gauge field
$A_{\mu}(x) \equiv \sum_a T_a A^a_\nu(x)$.  $S(A)^\dagger(x,y)$ denotes the
Hermitian adjoint of $S(A)(x,y)$ in color and Dirac space.
We use the relation
\beq
\gamma_5 S(A)(y,x)\gamma_5= S(A)^\dagger(x,y) \ . 
\label{gamma5}
\eeq
This property is unique to $\gamma_5$ and is not shared by any of the other 16
Dirac matrices. It ensures that the path integrand is positive for all field
configurations, making $P(x,y)$ larger than all other Euclidean (scalar,
vector, axial-vector, and tensor) correlators.  Asymptotically, when $|x-y| \to
\infty$, any correlator $C(x,y)$ behaves, up to a power-law prefactor, as
\beq
C(x,y) \simeq \exp(-m_0|x-y|)
\label{cxyasymptotic}
\eeq
where $m_0$ is the mass of the lightest physical state
with the quantum numbers of the correlator considered. This, together with the
inequality
\beq
P(x,y) \sim \exp(-m_{\pi}|x-y|) \ge \ \ {\rm any} \ \ C(x,y)
\label{pmpi}
\eeq
guarantees that the pion is, indeed, the lightest meson. Furthermore, the
positivity of $P(x,y)$ for any $|x-y|$ and the fact that $S(x,y)$ is a
monotonically decreasing function of $|x-y|$ implies that the
effective quark-antiquark potential in the pion (to the extent that this
nonrelativistic language is appropriate) is attractive at all relative
distances.

   As we noted in the previous section, in the nonrelativistic quark model a
very strong hyperfine interaction between the quark and antiquark in the pion
is needed in order to reduce its mass nearly to zero, and such an interaction
tends to produce a wavefunction for the valence $q \bar q$ in the pion that is
restricted to a very small spatial extent (almost collapsed).  Following this
lead, we suggest that while the spacetime (or Euclidean) picture of a $\bar q
q$ vector meson is two wool-ball-like single strands of valence quark and
anti-quark lines, the pion is a double strand, namely closer valence $\bar q$
and $q$ world lines whose motion forms a single wool-ball-like configuration.
According to this picture, in the pion, but not in the $\rho$ etc., the valence
$\bar{q}$ and $q$ lines with collinear momenta track each other at a distance
that is shorter than 1 fm. This is similar to NJL-type models, in which, by
construction, the important interaction is of short range.  Thus, in our
picture the pion qualitatively differs from the $\rho$ as a nearly
Nambu-Goldstone particle should and, at the same time, can be consistently
considered as a $\bar q_i q_j$ state. Here and below, analogous comments, with
obvious changes for the heavier $m_s$, apply for the $K$ and its comparison
with the $K^*$.

It is well known from discussions of the chiral anomaly
\cite{cw}-\cite{colemangrossman} that a massless collinear quark and antiquark
of opposite helicity, correponding to the bilinear operator product
$\bar\psi\gamma_5\psi$, can mimic the pole of a massless pseudoscalar particle
and replace the latter in the calculation of the anomaly.  Here we suggest that
such a configuration, including the effect of spontaneous chiral symmetry
breaking, can represent the massless pion, explain the puzzling strong color
hyperfine interaction between the $q_i$ and $\bar q_j$, and can account for its
behavior as a light, approximate Nambu-Goldstone boson. In general, one would
expect from the basic quantum mechanical relation $(\Delta p_i )(\Delta r_i)
\gsim \hbar$ that restricting the $q$ and $\bar q$ to a small interval along
some axis $\hat x_i$ would entail large momenta along this axis.  However, in
the presence of an appropriate gluonic field configuration, the
gauge-covariant momentum $p_\mu{\mathbb I} - gT^a A^a_\mu$ can vanish.

   We address the questions posed above, starting with (i). There are two
sources of explicit chiral symmetry breaking, namely finite quark masses and
the presence of nonzero electroweak interactions.  For our present discussion
we shall imagine that, unless otherwise indicated, electroweak interactions are
turned off.  Then a non-zero pion mass is induced via the explicit chiral
symmetry breaking term $m_u\bar{d}d+m_u\bar{u}u$ in the QCD Lagrangian.  To see
this in our picture, we consider the spacetime evolution of the $q_i$ and $\bar
q_j$ in a pion, after a Euclidean Wick rotation.  In the QCD context, the $q_i$
and $\bar q_j$ are connected by a chromoelectric flux tube, and their positions
fluctuate within length scales of order $1/\Lambda_{QCD} \simeq 1$ fm.  We
consider a model in which in the pion, but not in the $\rho$ or other vector
mesons, the valence $q$ and $\bar q$ track each other at a distance $h$ shorter
than $1/\Lambda_{QCD}$, at all times.  The constituent quark masses are then
less relevant to the dynamics, being gradually replaced, as $h$ gets smaller,
by their current-quark masses.  The point here is that constituent quark
masses are consequences of spontaneous chiral symmetry breaking, which
disappears at short distances (large momenta).  In QCD, the scale-dependent
dynamically generated constituenet quark mass decays, as a function of
Euclidean momenta $p$, like
\beq
M_q \sim \frac{\langle \bar q q \rangle}{p^2}
\label{mqdecay}
\eeq
up to logs \cite{sdeq}, where
\beq
\langle \bar q q \rangle \sim 4\pi f_\pi^3 \sim \Lambda_{QCD}^3 \ . 
\label{qqbarcondensate}
\eeq
More generally,
\beq
M_q \sim \Lambda_{QCD} \left ( \frac{\Lambda_{QCD}}{p} \right )^{2-\gamma} \ , 
\label{mqgen}
\eeq
where $\gamma$ denotes the anomalous dimension of the $\bar q q$ operator; here
we use the property that $\gamma$ is a power series in the running coupling
$\alpha_s$, and $\alpha_a$ approaches zero at short distances because of the
asymptotic freedom of QCD.  In our picture it is this ``melting away'' of the
constituent quark masses at short distance which provides, in the NQM language,
the very strong hyperfine interactions in the pion.

   Next, as an answer to question (ii), we would like to show how the GMOR
relation for the pion and other pseudoscalar meson masses (aside from the
$\eta'$), which embodies the Nambu-Goldstone nature of these pseudoscalar
mesons, is naturally expected in our picture.  Let the total Euclidean length
$R = |x|=\sqrt{\tau^2+{\bf r}^2}$, where $\tau = it$, be the net Euclidean
distance travelled by the $q \bar q$ double line describing the valence
quark-antiquark in the pion.  The double line describes a random walk with $n$
straight sections of total length $L$.  When probed at distances that are short
compared with $1/\Lambda_{QCD}$, the quark masses are the hard, current-quark
masses, $m_u \simeq 4$ MeV and $m_d \simeq 8$ MeV.  When $(m_u+m_d)|x| \gsim
1$, the quark propagation involves the suppression factor $e^{-(m_u+m_d)|x|}$.
Hence, a characteristic length describing this propagation is
\beq
L \propto \frac{1}{m_u+m_d} \ . 
\label{ell}
\eeq
Now the end-to-end distance $R$ for a random walk with step sizes $d$ (in any
dimension) is given by \cite{randomwalk}
\beq
R^2 \sim d^2 \, n \ . 
\label{rwrel}
\eeq
On average, if the total length of the $n$-step walk is $L$ and the step
length is $d$, then
\beq
d \simeq \frac{L}{n} \ . 
\label{drel}
\eeq
Hence,
\beq
R^2 \simeq Ld \ . 
\label{rld}
\eeq
Then the basic correlation function relation, Eq. (\ref{cxyasymptotic}) implies
that
\beq
m_\pi \simeq \frac{1}{R}
\label{mpir}
\eeq
and hence that
\beq
m_\pi^2 \simeq \frac{m_u+m_d}{d} \ . 
\label{mpirel}
\eeq
Since the step size $d$ is connected, via the helicity reversal process, to the
underlying confinement and dynamical breaking of chiral symmetry, it
is natural to equate 
\beq
\frac{1}{d} = -\frac{\langle \bar q q \rangle}{f_{\pi}^2} \ . 
\label{1overd}
\eeq
(where we follow the usual phase convention for the quark fields so that, with
$m_q$ taken as positive, the condensate $\langle \bar q q \rangle < 0$).
Combining these with Eq. (\ref{mpirel}), we see that this heuristic analysis
yields the GMOR mass relation,
\beq
m_{\pi}^2 = -\frac{(m_u+m_d)}{f_\pi^2} \, \langle \bar q q \rangle \ , 
\label{gmord}
\eeq
where $\langle q q \rangle \equiv \langle \sum_{a=1}^{N_c} \bar q_a q^a\rangle$
with $q=u$ or $q=d$ (these condensates being essentially equal in QCD).  A
similar argument, with appropriate replacement of light quark mass $m_u$ or
$m_d$ by $m_s$, yields the analogous GMOR-type mass relations for the $K^+$ and
$K^0$,
\beq
m_{K^+}^2 = -\frac{(m_u+m_s)}{f_K^2} \, \langle \bar q q \rangle
\label{mkprel}
\eeq
and
\beq
m_{K^0}^2 = -\frac{(m_d+m_s)}{f_K^2} \, \langle \bar q q \rangle \ , 
\label{mkzrel}
\eeq
where $\langle \bar q q \rangle \simeq \langle \bar s s \rangle$ for $q=u,d$.

The nearby $q$ and $\bar q$ paths in our picture generate $q$--$\bar q$ color
interactions that depend on the difference of these paths.  This suggests the
possibility of using this picture to infer the derivatively coupled form of
pion interactions appropriate for a Nambu-Goldstone particle.  This derivative
coupling means that in the static limit, these Nambu-Goldstone bosons become
non-interacting.

We next address point (iii) above, concerning the relation $f_\pi \sim
|\psi(0)|/\sqrt{m_\pi}$.  Since the matrix element (\ref{matrixelement}) as it
enters in the $\pi^+ \to \ell^+ \nu_\ell$ decay amplitude obviously involves
the annihilation of the $u$ and $\bar d$ quarks in the $\pi^+$ to produce the
virtual timelike $W^+$ that, in turn, produces the $\ell^+\nu_\ell$ pair, it
clearly depends on the $u \bar d$ wavefunction in the pion evaluated at the
origin of the relative coordinate, $|\psi_\pi(0)|$.  The question here concerns
what happens in the chiral limit, where $m_\pi \to 0$. For this discussion we
again imagine that electroweak interactions are turned off, except that we take
into account the couplings leading to the $\pi^+_{\ell 2}$ decay.  Now the pion
wavefunction at a given time involves the intersection of the worldlines of its
constituent $q$ and $\bar q$ with the $t=0$ hyperplane in the full ${\mathbb
  R}^4$ Wick-rotated spacetime.  This wavefunction has many Fock space
components.  The matrix element (\ref{matrixelement}) involves the annihilation
of the valence $q_i \bar q_j$ component by the axial-vector current. Higher
Fock space components in the pion wavefunction correspond to additional
crossings of the $t=0$ hyperplane.  A measure of the contributions of these
additional components can be obtained from our random walk representation.  We
note that for the present purpose it is essentially a one-dimensional random
walk that is relevant, since we are inquiring about passages across a
hyperplane, namely that defined by the condition $t=0$, of codimension 1 in the
full Euclidean ${\mathbb R}^4$.  Now in general, the number of times that a
one-dimensional random walk with $n$ steps returns to the origin is
asymptotically $\propto \sqrt{n}$ for large $n$. The contribution of the
valence $q \bar q$ component of the full pion wavefunction to the annhilation
probability $|\psi_\pi(0)|^2$ is thus reduced by the factor $1/\sqrt{n}$.  By
Eq. (\ref{rwrel}), $n^{-1/2} \propto R^{-1}$ and by Eq. (\ref{mpir}), $R^{-1}
\simeq m_\pi$, so $|\psi_\pi(0)|^2$ is reduced by the factor $m_\pi$.  This
means that $|\psi_\pi(0)| \propto \sqrt{m_\pi}$ in the chiral limit, thereby
cancelling the $\sqrt{m_\pi}$ in the denominator of Eq. ({\ref{vrw}), and
  yielding a finite value of $f_\pi$.  Similar remarks apply for $f_K$ in the
  hypothetical limit of $m_s \to 0$ as well as $m_{u,d} \to 0$. Thus, our
  picture provides a plausible resolution of the van Royen and Weisskopf
  paradox (Eq. (\ref{vrw})) \cite{vrw}.

Finally, we address issue (iv) concerning the similar size of the $\pi$ and
$\rho$. We should emphasize from the very outset that this is challenging.  The
qualitatively different physical pictures involved give an indication of the
complexity in the calculation of charge radii.  On the one hand, if the size is
controlled by the relatively small separation $h$ in our picture with double
$q_i \bar q_j$ lines, then the pion should be much smaller than the
$\rho$. On the other hand, since the distance $R \simeq 1/{m_\pi}$ controls the
overall pion size, it follows this size can become, at least formally,
unbounded in the chiral limit $m_\pi \to 0$. (In practice, pion wavefunctions
centered within a distance $R$ of each other would overlap and become
entangled.) This divergence in $R$ as $m_\pi \to 0$ is not an artifact of our
picture; the range of the residual strong force mediated, at long distance, by
pion exchange, formally diverges in this chiral limit. The property that the
pion charge radius also diverges in the chiral limit is a natural concomitant
of this divergence in the pion size.  Let us elaborate on this.

 The charge radius (squared) of a hadron is
\beq
\langle r^2 \rangle = \int \rho(r) \, r^2 \, d^3r \ , 
\label{rsqdef}
\eeq
where $\rho(r)$ denotes the charge density. The quantity $\sqrt{|\langle r^2
\rangle | }$ gives one measure of the size of a composite particle
\cite{rsqsize}. This is especially clear for a meson such as the $\pi^+$ or
$K^+$, where the $u$ and, respectively, $\bar d$ or $\bar s$ both contribute
positively to the integrand in Eq. (\ref{rsqdef}) \cite{rsqsize,size2}.

The charge radius squared is proportional to the slope of the electromagnetic
form factor $F(q^2)$ at $q^2=0$ \cite{r2ell}
\beq
\langle r^2 \rangle = 6 \, \frac{dF(q^2)}{dq^2}{} \Big |_{q^2=0} \ . 
\label{dfdqsq}
\eeq

The latter form factor satisfies a $t$-channel dispersion relation ($t \equiv
q^2$)
\beq
F(t) = \int dt \, \frac{{\rm Im}[F(t')]}{t-t'} \ . 
\label{disprel}
\eeq
In particular, for the case under consideration, $F(t) = F_{\pi^+}(t)$, the
integration is from $t'=(2m_\pi)^2$ to $t'=\infty$. In the vector meson
dominance approximation for $F(t)$, one commonly replaces the ${\rm Im}[F(t')]$
by a delta function corresponding to the approximation of zero-width for the
relevant vector meson.  Here, using $\rho$-dominance for $F_{\pi^+}(t)$, one
replaces ${\rm Im}[F_{\pi^+}(t')]$ by a delta function $\propto
\delta(t'-m_\rho^2)$.  This narrow-width approximation, together with the known
value $F_{\pi^+}(0)=1$, yields
\beq
F_{\pi^+}(q^2) = \frac{m_\rho^2}{m_\rho^2-q^2} \ , 
\label{fpivdm}
\eeq
so that
\beq
\sqrt{\langle r^2 \rangle_{\pi^+}} = \frac{\sqrt{6}}{m_\rho} = 0.62 \ {\rm fm}
\ . 
\label{rsqpivdm}
\eeq
This is close to the experimentally measured value, given above in Eq.
(\ref{rsqpip}) \cite{bauerpipkin,gebrown}; quantitatively, it is smaller than
this experimental value by only 7 \%.  (One can also include the effects of the
$\rho$ width, but this will not be necessary for our discussion here.) An
analogous vector meson dominance prediction for the $K^+$ charge radius works
very well also \cite{bauerpipkin}.  {\it A priori}, one might worry that an
additional threshold contribution from $t' \simeq (2m_{\pi})^2$ might dominate
and lead to $\langle r^2 \rangle_{\pi^+} \simeq 1/m_\pi$.  However, this does
not happen here because of the derivative coupling of soft pions, as
Nambu-Goldstone bosons. In the particular case here, another reason why this
does not happen is that there is a $\sqrt{t'-4m_{\pi}^2}$ factor in ${\rm
Im}(F(t'))$ that arises from the $P$-wave nature of the $\pi \pi$ amplitude.
Nevertheless, $\langle r^2 \rangle_{\pi^+}$ does diverge as $\langle r^2
\rangle_{\pi^+} \sim \ln(1/m_\pi)$ in the chiral limit where $m_\pi \to 0$
\cite{gl}.

We next sketch an estimate of the pion charge radius in our picture. As
in the previous section, the higher Fock space states play a key role in this
estimate.  Consider the $t=0$ slice of the Wick-rotated Minkowski space.  The
quantity $\langle r^2 \rangle_{\pi^+}$ can be computed as a sum of the
contributions of the various Fock space components of the pion wavefunction.
In our picture these are generated by crossings of the $t=0$ hyperplane by the
$q_i \bar q_j$ random-walking double worldline of the pion.  Let the $k$'th
such crossing be at ${\bf r}_k$. The first crossing corresponds to a $\pi^+$,
say, moving forward in time. Hence, we have a charge $+1$ at this location. At
the second crossing, the pion line is reversed, and we have a $-1$ charge at
${\bf r}_2$, etc. The definition Eq. (\ref{rsqdef}) above then yields $\langle
r^2 \rangle_{\pi^+} \simeq \sum_{k=1}^\infty (-1)^k r_k^2$, where here $r_k
\equiv |{\bf r}_k|$. Recalling that the $k$'th visit to the $t=0$ plane happens
typically after $n \simeq k^2$ steps of the random-walking double line, with
individual step size $d$, we deduce that, on average, $r_k^2 \simeq k^2 d^2$.
By itself, this would yield, for $\langle r^2 \rangle_{\pi^+}$, the sum
$\sum_{k=1}^\infty (-1)^k k^2 d^2$.  The terms in the above oscillating sum
diverge as $k \to \infty$.  However, to get the actual sum, we must take into
account the fact that the contributions are regularized by the exponential
$\exp[-(m_u+m_d)L] \simeq \exp[-(m_u+m_d)k^2d]$ controlling the total length of
the random-walking double line. Using $m_u+m_d = d m_{\pi}^2$ from
Eq. (\ref{mpirel}) above and defining
\beq
b \equiv m_{\pi} d \ , 
\label{bdef}
\eeq
we can rewrite the charge radius as
\beq
\langle r^2 \rangle_{\pi^+} \simeq \sum_{k=1}^\infty  \, 
(-1)^k k^2 d^2 e^{-k^2b^2} \ . 
\label{sum3}
\eeq
Since $k$ gets very large in the chiral limit, there are strong
cancellations between successive terms, rendering an accurate estimate
difficult. We can at least investigate the nature of the leading divergence in
$\langle r^2 \rangle_{\pi^+}$.  To do this, we replace the above sum, after
subtracting and adding an $r_0$ term and symmetrizing, by an integral over the
variable $\xi =k b = k m_\pi d$:
\beqs
I(b) & = & \frac{1}{m_\pi^2} \, \int_{-\infty}^{\infty} d \xi \ 
\xi^2 \, \exp \left ( \frac{i\pi \xi}{b} - \xi^2 \right ) \cr\cr
& & \cr\cr
&=& \frac{\sqrt{\pi}}{2m_\pi^2} \, \left (1-\frac{\pi^2}{2(m_\pi d)^2}\right ) 
\exp \left [ -\left (\frac{\pi}{2m_\pi d} \right )^2 \right ]
\cr\cr
& & 
\label{int3}
\eeqs
The key observation here is that, while we have, as expected, an explicit
$1/m_\pi^2$ factor in front, the integral $I(b)$ and any finite derivative
thereof, contain the factor $\exp[-\pi^2/(2m_\pi d)^2]$, which vanishes with an
essential zero in the chiral limit $m_\pi \to 0$.  This can be seen as a
consequence of the strong cancellations between different terms contributing to
the sum, which we approximated as an integral.  Thus, our calculation shows the
absence of a divergence of the power-law form $1/m_{\pi^+}^2$ in $\langle
r^2\rangle_{\pi^+}$ and is consistent with the chiral perturbation theory
result that $\langle r^2\rangle_{\pi^+}$ diverges like $\ln(1/m_\pi)$ in this
limit.  For the real world with nonzero current quark masses for $u$ and $d$,
our analysis above naturally yields a value of $\langle r^2\rangle_{\pi^+} \sim
d^2$, since this was the $r_0$ term in the sum. A similar conclusion, with
appropriate replacement of the $d$ with the $s$ quark, applies to $\langle r^2
\rangle_{K^+}$ in the SU(3) chiral limit $m_u, \ m_d, \ m_s \to 0$.

Our picture can also give a plausible explanation of why the pion-nucleon cross
section $\sigma_{\pi N}$ at energies above the resonance region can be
comparable to the inferred value of $\sigma_{\rho N}$ at the same energies
(c.f. Eqs. (\ref{sigma_pin}) and (\ref{sigma_rhon})).  Relevant to the $\pi N$
cross section is the fact that the valence quarks in the pion propagate in an
extended double-line manner covering an area of order $R^2 \sim 1/m_\pi^2$.
However, because of the strong color hyperfine interaction, the separation $h$
of the valence $q_i$ and $\bar q_j$ in the pion is rather small in our model.
Hence, while in a crossing of two $q_i \bar q_j$ pairs at an ordinary hadronic
distance $\sim d$, the probability of an interaction is O(1), here, in
contrast, it will be $O((h/d)^2)$.  In the context of a hadronic string
picture, the small pion mass is related to the separation $h$ via $m_\pi
\propto \sigma h$, where $\sigma$ is the the hadronic string tension. Hence,
one may roughly estimate that the $\pi N$ cross section $\sigma_{\pi N}$
contains the factors $(\pi R^2) (h/d)^2 \propto \pi/(\sigma d)^2$. Note that
the factor of $m_\pi^2$ cancels out between numerator and denomator, leaving
$\sigma_{\pi N}$ proportional to an expression involving the string tension and
a typical hadronic distance scale, which are the same for the $\pi$ and the
$\rho$.

 In the preceeding we have presented our efforts to show how our picture of a
rather tightly bound $q_i \bar q_j$ pair undergoing a random walk inside a pion
can explain how this particle can exhibit the properties of an approximate
Nambu-Goldstone boson while also being understandable as a $q \bar q$ bound
state.  Ultimately, one should be able to find the differences predicted by our
picture as compared with other approaches to this physics. One theoretical tool
that is relevant here is lattice gauge theory.  However, one faces not only the
technical difficulty of simulating very light quark masses and light pions.  An
additional challenge is that in (Euclidean) lattice simulations one first
integrates over the fermionic degrees of freedom. Having the two (say $u$ and
$\bar d$) quark propagators in the same background color field may not allow
one to verify that at all intermediate steps the quark and antiquark are really
close to each other. One may need to go back to the sum over fermionic paths in
order to actually detect the propagators of the nearby $q_i \bar q_j$ pair. 

One implication of our model with the nearby $q_i \bar q_j$ lines separated by
a relatively small distance $h$ is that the purely gluonic exhange amplitude
for $\pi \pi$ scattering should be rather small.  A recent lattice calculation
of the $I=2$ $\pi \pi$ $S$-wave scattering length obtained the result $a_2
\simeq -0.043/m_\pi$ \cite{bedaq}, in agreement with the Weinberg-Tomozawa
soft-pion current algebra result $a_2 = -m_\pi/(16 \pi f_\pi^2) = -0.044/m_\pi$
\cite{weintom,ad}.  In a hypothetical $\pi \pi'$ scattering, where the $\pi'$
is comprised of $\bar d'$ and $u'$ quarks that are degenerate with the ordinary
$u$ and $d$ but do not mix with them, the scattering amplitude involves only
gluon exchanges, but not quark interchanges.  In this case a preliminary
lattice calculation has obtained a $\pi \pi'$ scattering length considerably
smaller than $a_2$, in qualitative agreement with our discussion above
\cite{walk}.

\section{Some Comments on the $K \to \pi$ and Heavy Quark Transition Form
Factors}

   The $K$ mesons undergo semileptonic $K_{\ell 3}$ decays, such as $K^+ \to
\pi^0 \ell^+ \nu_\ell$, $K^0_L \to \pi^+\ell^-\bar\nu_\ell$, and $K^0_L \to
\pi^- \ell^+\nu_\ell$, mediated by the vector part of the weak charged
current. The almost conserved vector current (CVC) (conserved apart from SU(3)
flavor-breaking effects) helps to fix the corresponding hadronic matrix
elements and the Cabibbo-Kobayashi-Maskawa (CKM) quark mixing matrix element
$|V_{us}|$ \cite{ckm} and tests of three-generation CKM unitarity
\cite{cabfit}. Denoting the 4-momenta $p=p_K+p_\pi$ and $q=p_K-p_\pi$,
one has
\beqs
     \langle \pi^0|(V_-)_\lambda|K^+ \rangle  & = &
     \langle \pi^-|(V_-)_\lambda|K^0_L \rangle = \cr\cr
& & \frac{1}{\sqrt{2}}(f_+(q^2)p_\lambda + f_-(q^2)q_\lambda) \ , 
\label{amp}
\eeqs
where $(V_-)_\lambda$ is the weak charged current. The contribution from the
$f_-(q^2)$ term is proportional to the final lepton mass squared and is
negligible for semileptonic decays to final electrons. The $t \equiv q^2$
variation of $f_+(t)$ over the range $m_e^2 \le t \le (m_K-m_{\pi})^2$ can be
approximated by a linear function of $t$:
\beq
f_+(t) = f_+(0)\left ( 1 + \lambda_+ \frac{t}{m_\pi^2} \right ) \ , 
\label{lambdadef}
\eeq
where $\lambda_+= 0.0288$, \cite{pdg}, in agreement with chiral perturbation
theory calculations \cite{gl,chipt} and also with the expectation from simple
$K^*$ vector meson dominance. CVC implies that $f_+(q=0)=1$ in the hypothetical
limit of exact SU(3)$_V$ symmetry, $m_u=m_d=m_s$ and hence $m_K = m_\pi$.  A
general result which we will elaborate on later is that the corrections to this
symmetry-limit value are always second-order in SU(3)$_V$ breaking.  This is
the well-known Ademollo-Gatto theorem \cite{ag}.  This feature is evident in
the explicit estimate \cite{lp,wada}
\beq
   f_+(0)=1-\frac{5(m_K^2-m_\pi^2)^2}
           {384\pi^2 f_\pi^2 (2m_K^2 + m_\pi^2)} =  0.985
\label{lpren}
\eeq
The correction term in Eq. (\ref{lpren}) arises from multi-particle
contributions to the sum-rule corresponding to the commutator $[Q_{4+i5},
Q_{4-i5}] = Q_3 + \sqrt{3}Q_8$, where the subscripts refer to SU(3) flavor
generators.  Formally, this sum rule underlies the Ademollo-Gatto theorem; the
multi-particle contributions are squares of matrix elements of the divergence
$\partial_\mu J^\mu_{4+i5}$ of the strangeness-changing weak vector current,
making the deviation from universality quadratic in the SU(3)$_V$ symmetry
breaking.  However, as is also evident in (\ref{lpren}), in the limit of ${\rm
SU}(3)_L \times {\rm SU}(3)_R$ chiral symmetry, with $m_\pi \to 0$ and $m_K \to
0$, the correction term is actually of first-order \cite{lp}.  This is a
consequence of the fact that in this chiral limit the contributions to the sum
rule that involve the exchange and propagation of massless $\pi$'s and $K$'s
lead to a deviation from universality that is linear in $m_K^2$.  The
correction in Eq. (\ref{lpren}) is small partly because of the numerical
coefficient arising from the loop diagram involved in the calculation.

It is instructive to see how the Ademollo-Gatto theorem is realized in the NQM,
where the form factor $f_+(q^2)$ can be expressed as an overlap of the $K^+$
and $\pi^+$ wavefunctions, which we shall denote $F_{K \to \pi}(q^2)$. In the
NQM, the $\pi^+$ and $K^+$ consist of nonrelativistic constituent quarks $q_i
\bar q_j$ and for $L=0$, and:
\beq
     F_{K \to \pi}(q=0) = \int d^3 r \psi_\pi({\bf r})^* \psi_K({\bf r}) \ . 
\label{fk}
\eeq
In this model the $K$ and $\pi$ wavefunctions (which are real) depend on just
an overall flavor independent potential $V(r)$ and on the reduced masses
$\mu_K$ and $\mu_\pi$.  We recall our notation $M_q$ for the constituent mass
of a quark $q$, and the values $M_u=M_d \equiv M_{ud} \simeq 330$ MeV,
$M_s \simeq 470$ MeV.

   For simplicity we use the single-term form for the potential:
\beq
V_{q \bar q}(r) = V_0 \left ( \frac{r}{r_0} \right )^\nu \ , 
\label{v}
\eeq
where $\nu$ is an exponent.  Special cases include (i) $\nu=-1$, i.e.,
Coulombic, (ii) $\nu=0$, with $V(r) \propto \ln r$; (iii) $\nu=1$, linear; (iv)
$\nu=2$, harmonic oscillator; and (v) $\nu = \infty$, equivalent to an infinite
square-well (ISW) potential. As was noted above, a realistic quark-quark
potential has different forms at short distances and at distances of order
$1/\Lambda_{QCD} \sim 1$ fm, so it is more complicated than a single-term form.
However, the simplification will suffice for our purposes here.  The scaling
properties of the Schr\"odinger equation imply that the spatial extent $r$
characterizing the falloff of the wavefunction scales with the reduced mass
$\mu$ as \cite{qr,vform}
\beq
r \propto \mu^{-\frac{1}{2+\nu}} \ . 
\label{psi_scaling}
\eeq
For the range of $\mu$ considered here, the dependence of this characteristic
distance on $\nu$ is thus maximal for the Coulombic, $\nu=-1$, case and minimal
for $\nu=\infty$, where the spatial extent of $\psi$ is determined completely
by the width of the infinite square well and is independent of $\mu$. 

For $\nu=2$, i.e., the harmonic oscillator potential, which we write as $V = k
r^2/2$, the wavefunction is proportional to a Hermite polynomial, and, for
the ground state, it is
\beq
\psi = \frac{(\mu k)^{3/8}}{\pi^{3/2}}\exp \left (
   \frac{\sqrt{\mu k} \, r^2}{2} \right ) \ . 
\label{psi_oscillator}
\eeq
Substituting this into Eq. (\ref{fk}), we calculate
\beq
F_{K \to \pi}(0) = \frac{2^{3/2}(\mu_K\mu_\pi)^{3/8}}
{(\sqrt{\mu_K}+\sqrt{\mu_\pi} \ )^{3/2}} \ . 
\label{fkp_oscillator}
\eeq
Let us define the following measure of flavor SU(3) symmetry breaking:
\beq
\epsilon = \frac{M_s-M_{ud}}{M_{ud}} \ . 
\label{eps}
\eeq
The expression for $F_{K \to \pi}(0)$ in Eq. (\ref{fkp_oscillator}) has the
following Taylor series expansion in $\epsilon$:
\beq
F_{K \to \pi}(0) = 1 - \frac{3}{256}\epsilon^2 + O(\epsilon^3)  
\quad {\rm for} \ \ \nu=2
\label{fkp_oscillator_taylor}
\eeq
With the values of $\mu_\pi$ and $\mu_K$ given above,
\beq
F_{K \to \pi}(0) = 0.999 \quad {\rm for} \ \ \nu=2 \ . 
\label{f0nu2}
\eeq
For comparison, consider the Coulomb potential with ground-state
\beq \psi=\frac{e^{-r/a}}{\pi^{1/2}a^{3/2}}
\label{psi_coul}
\eeq
where
\beq
a_B = \frac{1}{C_{2f}\alpha_s \mu} 
\label{abohr}
\eeq
is the Bohr radius and $C_{2f}=4/3$. Substituting this into Eq. (\ref{fk}) for
the wavefunction overlap, we find
\beqs
      F_{K \to \pi}(0) & = &
\left ( \frac{2\sqrt{a_K a_\pi}}{a_K+a_\pi} \ \right )^3 =
\left (\frac{2\sqrt{\mu_K \mu_\pi}}{\mu_K + \mu_\pi} \ \right )^3 \cr\cr
& = & 0.992
\label{fkp_coulomb}
\eeqs
This again has a Taylor series expansion of the form $F_{K \to
\pi}(0)=1-O(\epsilon^2)$, as expected from the Ademollo-Gatto theorem.  Since
the $r$-dependences of the logarithmic ($\nu=0$) and linear ($\nu=1$)
potentials are intermediate between the harmonic oscillator ($\nu=2$) and
Coulomb ($\nu=-1$) potentials, one expects $F(q=0)$ to be very close to unity
for these potentials as well.

These results do not imply such small deviations from unity for the form factor
$f_+(0)$ in $K_{e3}$ decay.  The mass difference $m_K-m_\pi \sim 360$ MeV far
exceeds the value of $m_s-m_u$ expected in a model with a flavor-independent
confining potential.  Such considerations would apply better to semileptonic $s
\to u$ decays of mesons with heavy $c$ or $b$ spectator quarks.  Indeed
$m_{D_s}-m_{D_u} = 104$ MeV and $m_{B_s}-m_{B_u} = 89$ MeV, consistent with the
current quark mass difference $m_s-m_u$. Unfortunately, these small mass
differences imply tiny branching for these decays $D_s \to D_u \ell^+ \nu_\ell$
and $B_s \to B_u \ell^+ \nu_\ell$.

   The generic form factor is a Lorentz-invariant function of $q^2$.  However,
for elastic scattering, one can go to a frame where $q^0=0$ so that $q^2 =
-|{\bf q}|^2$ and write
\beq
   F(q^2) = \int d^3 r e^{i {\bf q} \cdot {\bf r}} \psi_\pi({\bf r})^*
\psi_K({\bf r}) \ , 
\label{formfactor}
\eeq
where $q=(q^0,{\bf q})$ is the momentum imparted to the leptons in the decay
process, ${\bf r} = {\bf r}_q - {\bf r}_{\bar q}$, and $\psi_K$ and $\psi_\pi$
are the initial and final meson wavefunctions.  In the flavor SU(3) symmetry
limit, $\psi_K=\psi_\pi$. For $q=0$, the normalizations of the wave functions
imply the conserved vector current (CVC) value $F(0)=1$.

The last result is quite general; if the mesons contain, in addition to the
valence quarks $q_i$ and $\bar q_j$, any number of gluons at the position
${\bf R}_s$ and/or $q \bar q$ quark pairs at the positions ${\bf r}_\ell$,
${\bf r}_{\ell'}$, we would have, instead of (\ref{formfactor}), 
\begin{widetext}
\beq
F_{K \to \pi}(q^2) = \int d^3 r \, e^{i {\bf q} \cdot {\bf r}} \Big [ 
   \prod_{\ell,\ell',s} \, d^3r_\ell \, d^3 r_{\bar\ell'} \, d^3 R_s \ 
 \psi_\pi({\bf r},{\bf r}_\ell,{\bf r}_{\bar\ell'},{\bf R}_s)^* 
    \psi_K({\bf r},{\bf r}_\ell,{\bf r}_{\bar\ell'},{\bf R}_s) \Big ] \ , 
\label{formfactorgen}
\eeq
\end{widetext}
so that again in the flavor SU(3) symmetry limit, for equal wavefunctions and
$q=0$, we have $F(0)=1$.  Here, $\psi_K$ and $\psi_\pi$ are the Fock space
wavefunctions with any number of gluons and quark-antiquark pairs. Both quarks
and gluons carry spin and color, so that $\psi$ could be a superposition of
many color and spin couplings which yield overall color singlets.  For
notational simplicity we have omitted these above. The general arguments
presented below do not depend on the slightly simpler form of
(\ref{formfactorgen}).

As is evident in eqs. (\ref{formfactor}) and (\ref{formfactorgen}), deviations
from $F(0)=1$ can be caused in two ways.  First, even for elastic transitions
with $\psi_{initial}=\psi_{final}$, the momentum transfer factor $e^{i {\bf q}
\cdot {\bf r}}$ modulates the positive integrand and decreases $F$. Second,
flavor SU(3) breaking, namely the difference between $m_s$ and $m_q$, $q=u,d$,
causes the $\pi^+$ and $K^+$ wavefunctions to be different and hence reduces
$f_+(0)$ from unity.  To analyze this, we shall use the Cauchy-Schwarz
inequality, that for any vector space ${\cal V}$ with vectors $\psi$ and $\phi$
and an inner product $\langle \psi,\phi\rangle$, the property
\beq
|\langle \psi,\phi\rangle| \le \|\psi \| \, \| \phi \|
\label{schwarzgen}
\eeq
holds, where $\| \psi \| \equiv \sqrt{\langle \psi, \psi\rangle)}$.  We apply
this to the $L^2$ Hilbert space of square-integrable functions
$\psi({\bf r},{\bf r}_\ell,{\bf r}_{\bar\ell'},{\bf R}_s)$ with the inner
product
\begin{widetext}
\beq
\langle \psi,\phi \rangle = \int d^3 r \, \Big [ \prod_{\ell,\ell',s} \, 
d^3r_\ell \, d^3r_{\bar\ell'} \, d^3R_s \ 
\psi({\bf r},{\bf r}_\ell,{\bf r}_{\bar\ell'},{\bf R}_s)^*
\phi({\bf r},{\bf r}_\ell,{\bf r}_{\bar\ell'},{\bf R}_s) \Big ] \ . 
\label{innerprod}
\eeq
\end{widetext}
Thus,
\beq
F_{K \to \pi}({\bf q}=0) = \langle \psi_\pi, \psi_K \rangle \ . 
\label{finner}
\eeq
Using this, we have
\begin{widetext}
\beqs
& &     | F_{K \to \pi}({\bf q} =0)|^2 = 
\bigg | \int d^3r \, \Big [ \prod_{\ell,\ell',s} \, 
d^3r_\ell \, d^3r_{\ell'} \, d^3R_s 
  \ \psi_\pi({\bf r},{\bf r}_\ell,{\bf r}_{\bar\ell'},{\bf R}_s)^*
    \psi_K({\bf r},{\bf r}_\ell,{\bf r}_{\bar\ell'},{\bf R}_s) \Big ] 
\bigg |^2  \cr\cr
&\le & 
\bigg [ \int d^3r \, \Big [ \prod_{\ell,\ell',s} \,
d^3r_\ell \, d^3r_{\ell'} \, d^3R_s 
|\psi_\pi({\bf r},{\bf r}_\ell,{\bf r}_{\bar\ell'},{\bf R}_s)|^2 \Big ] \bigg ]
\bigg [ \int d^3r \, \Big [ \prod_{\ell,\ell',s} \, 
d^3r_\ell \, d^3r_{\ell'} \, d^3R_s 
|\psi_K({\bf r},{\bf r}_\ell,{\bf r}_{\bar\ell'},{\bf R}_s)|^2   \Big ] \bigg ]
\cr\cr
& & 
\label{schwarz}
\eeqs
\end{widetext}

where we write these for the general case of Eq. (\ref{formfactorgen}) above.

Universal form factors at the no-recoil point for semileptonic decays of mesons
containing heavy quarks, e.g., $B_d \to D^- \ell^+ \nu_\ell$, follow from the
fact that for $m_b > m_c >> \Lambda_{QCD}$, the heavy quark is a static source
of color (transforming as a color SU(3) triplet) with common wavefunctions for
all of the light degrees of freedom in either the $B$ or $D$ mesons
\cite{nusswetz}.  With flavor-independent primary QCD interactions, the
difference between $\psi_K$ and $\psi_\pi$ is due to the different $u$ and $s$
masses only.  From the Cauchy-Schwarz inequality one sees that $F({\bf q} =
{\bf 0})$, as a function of $m_s$ and $m_s-m_u$, is extremal (maximal) at
$m_s-m_u=0$. Hence the deviation from unity is of order $O((m_s-m_u)^2)$, which
is the Ademollo-Gatto theorem \cite{ag}. The analogous theorem for heavy quarks
is derived by the same type of reasoning \cite{luke,hqrev}. Let us rewrite
(\ref{schwarz}) as
\beq
   F_{B \to D}({\bf q}={\bf 0})[m_i(``light''),v=m_i/M_Q]
\label{lukeq}
\eeq
where $m_i(``light'')$ refers to the masses of the degrees of freedom that are
light relative to $m_Q$, namely $\Lambda_{QCD}$, $m_s$, etc. and $M_Q$ denotes
the mass of the lighter among the heavy quarks, namely $m_c$ in the present
case. Again, $F$ is extremal for $v=0$, and the deviations from universality at
the no-recoil point are of order $O(v^2)= O(1/m_Q^2)$, i.e., $O(1/m_c^2)$ in $b
\to c$ transitions.  Indeed, if the current quark masses $m_u=m_d=0$, then,
when $m_s \to 0$, the chiral symmetry group is enlarged from ${\rm SU}(2)_L
\times {\rm SU}(2)_R$ to ${\rm SU}(3)_L \times {\rm SU}(3)_R$ (and the QCD
condensates would then break these to the respective diagonal subgroups
SU(2)$_V$ and SU(3)$_V$).

The generalized Ademollo-Gatto theorem can be formulated in Hamiltonian lattice
QCD.  The $\pi$ and $K$ wavefunctions are replaced by wavefunctionals with
arbitrary patterns of excited links, corresponding to gluonic excitations,
and/or extra $q \bar q$ pairs. Now consider the matrix element of the
strangeness-changing vectorial weak charge, $Q_{u, \bar s}= \int d^3 x \, V^0$,
where $V^\mu$ denotes the associated current. This converts flavors $s \to u$
for the valence quarks. Since $[Q,H] \ne 0$, this changes the energy of the
state operated on by $m_K-m_\pi$.  However, since $Q$ is an integral over all
space, it does not change the 3-momentum of the state on which it
operates. Hence, if it operates on a $K$ at rest, it should produce a $\pi$ at
rest also.  The matrix element of interest is the overlap of two
wavefunctionals computed for valence quark mass $m_q=m_s$ and for $m_q=m_u$. By
the Cauchy-Schwarz inequality, which holds for these wavefunctionals, the
overlap is smaller than unity, achieving its maximum value when $\Delta=m_s-m_u
= 0$. Hence, repeating the same arguments as above, we find that
\beq
F({\bf q}={\bf 0})= 1 - O(\Delta^2) \ .
\eeq

\section{Mass Comparisons Involving Heavier Hadrons}

We proceed to discuss the systematics of mass differences $m(Q \bar s) - m(Q
\bar u)$ for various $J^{PC}$ mesons. Some related work is in Refs.
\cite{kl,kklr}. In this context, we recall that modern lattice estimates have
yielded a somewhat smaller value of the current quark mass $m_s \sim 120$ MeV
than some older current algebra estimates, which tended to be centered around
180 MeV \cite{gl}.  In the nonrelativistic quark model (with a
flavor-independent nonrelativistic quark-(anti)quark interaction potential),
the mass difference between analogous hadrons differing only by having an $s$
quark replaced by a $u$ or $d$ quark should, up to small binding changes due to
the different reduced constituent masses, differ by $m_s-m_{ud}$.  The real
world is more complicated, for several reasons. First, the concept of quark
masses and differences needs to be carefully defined.  The masses run with the
distance or momentum scale at which they are probed. The constituent quarks can
be considered to be extended quasiparticles, confined to hadrons with sizes of
order $1/\Lambda_{QCD}$.  As the MIT-SLAC deep inelastic scattering experiments
showed drmatically, as one increases the momentum scale at which one probes
such a quark beyond $\Lambda_{QCD}$, it acts quasi-free, without the attendant
strong coupling to gluons to which it is subject for momenta less than
$\Lambda_{QCD}$.  As this momentum scale increases considerably beyond
$\Lambda_{QCD}$, the quark mass then goes over to approximately the current
quark mass, since the QCD gauge coupling becomes small.  Since different
hadrons have somewhat different effective scales, this modifies the extracted
mass difference.

Secondly, while at the fundamental Lagrangian level the only breaking of flavor
symmetry is due to the differences between the current quark masses, this is
not the case for the effective potential between the constituent quarks in the
naive quark model because of the short-range color hyperfine interactions,
though not in the asymptotic, confining part of the potential.  This suggests
that the mass differences of $Q \bar s$ and $Q \bar q$ mesons with $Q$ a heavy
quark better estimate the current quark mass difference $m_s-m_q$ mass
difference, with $q=u$ or $d$, since both the magnitude of the color hyperfine
splittings and the effective sizes of the system are smaller there (the latter
is a reduced mass effect).  Some measured mass differences, averaged over
isospin multiplets, are $m(K^*)-m(\rho) \simeq 120$ MeV, $m(\phi)-m(K^*) \simeq
125$ MeV, $m(D_s)-m(D_u) \simeq m(D_s^*)-m(D_u^*) \simeq 100$ MeV, and
$m(B_s)-m(B_u) \simeq 90$ MeV.  We observe a substantial and fairly systematic
tendency of these mass difference to decrease as $m(Q)$ increases.  This is in
agreement with the lattice gauge theory estimates mentioned above. The pattern
in the baryonic spectrum is more complicated, but does not disagree with this
general decreasing behavior.  As is well known, the large splittings in the
$J^P=1/2+$ baryon octet, viz., $m(\Lambda)-m(N) \simeq 180$ MeV,
$m(\Sigma)-m(N) \simeq 255$ MeV, $m(\Xi) - m(\Lambda) \simeq 200$ MeV, and
$m(\Xi) - m(\Sigma) \simeq 125$ MeV, can be explained by a color hyperfine
interaction, similar to that for the mesons.  The equal-spacing mass difference
rule in the $J=3/2$ baryon decuplet with the interval of $\sim 146$ MeV can
also be explained by the color hyperfine interaction.  The relatively large
mass difference between $(csu,1/2+) \equiv \Xi_c$ and $(cud,1/2^+) \equiv
\Lambda_c$ of $m(\Xi_c)-m(\Lambda_c) \simeq 181$ MeV is again in agreement with
the expectation based on the large difference in the $s-u$ and $d-u$ color
hyperfine interaction, which is evidently not reduced by the presence of the
nearby heavy $c$ quark in these baryons.  Only the difference of masses of
$\Omega_c = (css,1/2+)$ and $\Xi_c = (csu,1/2+)$ of 230 MeV appears to be
somewhat high.  On the basis of this discussion, one expects small mass
splittings $m(QQ's)-m(QQ'u)$ between baryons containing two heavy quarks, but
this expectation cannot yet be checked.

\section{Conclusions}

In conclusion, we have revisited the $\rho$--$\pi$ puzzle, namely, the problem
of describing the $\pi$ meson as a $q_i \bar q_j$ bound state and as an
approximate Nambu-Goldstone boson and relating its mass and size to those of
the $\rho$ meson. We have presented a simple heuristic picture that, we
believe, gives insight into this problem. In this picture, the valence $q_i$
and $\bar q_j$ quarks in the $\pi$ are rather tightly bound by the strong color
hyperfine interaction that splits the $\pi$ and $\rho$ masses.  We show that
this picture can resolve another old puzzle concerning the pion wavefunction at
the origin (van Royen-Weisskopf paradox) and is consistent with the
Gell-Mann-Oakes-Renner relation, With appropriate replacement of the $u$ or $d$
quark by the $s$ quark, our picture also applies to the $K$ and its relation to
the $K^*$. Using our model, we present an estimate for the charge radius
$\langle r^2 \rangle_{\pi^+}$.  Our approach gives further insight into the
charged-current $K^+$ -- $\pi^+$ transition relevant in $K_{\ell 3}$ decays.

\begin{acknowledgments}

S. N. thanks A. Casher and M. Karliner, and R. S. thanks S. Brodsky, for 
helpful discussions. The research of R. S. was partially supported by the grant
NSF-PHY-06-53342.

\end{acknowledgments}

\vfill
\eject


\begin{thebibliography}{99}
\bibitem{qmass}
%
For the light quarks $u$ and $d$, the constituent quark mass $M_u \simeq M_d$
to within a few MeV, so we shall denote it as $M_{ud}$. The value of this mass
can be estimated roughly as $M_{ud}=m_N/N_c = 310$ MeV or $m_{ud}=m_\rho/2 =
380$ MeV; we shall essentially average these and use the value $M_{ud}=340$ MeV
here, as in Eq. (\ref{mud}). $M_{ud}$ and $M_s$ can be obtained by a fit to
hadron masses and baryon magnetic moments in the NQM.  The current-quark
masses, i.e., the masses that quarks would have in the hypothetical absence of
strong interactions, are denoted $m_i$; typical values for these are $m_u
\simeq 4$ MeV, $m_d \simeq 8$ MeV, and $m_s \simeq 120$ MeV \cite{pdg}. These
are often called hard quark masses, although, if they are dynamically
generated, they themselves are soft at mass scales typically of order $10^2$ 
to $10^3$ TeV \cite{sml}.
\bibitem{rsqsize}
%
Some illustrative semiclassical examples with total charge $Q=\int \rho(r) d^3r
= 1$ and positive-definite $\rho(r)$ which have constant, exponential, and
Gaussian forms, respectively, are the following: (i) if $\rho(r) = 3/(4 \pi
R^3)$ for $r \le R$ and $\rho(r) = 0$ for $r > R$, then $\langle r^2\rangle =
(3/5)R^2$; (ii) if $\rho(r) = (8 \pi R^3)^{-1}e^{-r/R}$, then $\langle r^2
\rangle = 12R^2$; and (iii) if $\rho(r) = (\sqrt{\pi} \, R)^{-3} e^{-(r/R)^2}$,
then $\langle r^2 \rangle = (3/2)R^2$.
\bibitem{size2}
%
The connection between $\sqrt{|\langle r^2 \rangle|}$ and the size of the
hadron is less direct for a particle in which the quarks contribute with
opposite signs to the integrand of Eq. (\ref{rsqdef}).  Indeed, because
$\langle r^2 \rangle$ flips sign under charge conjugation, $\langle r^2 \rangle
= 0$ for a self-conjugate particle, such as $\pi^0$.  However, isospin
invariance implies that the size of $\pi^0$ is equal, up to small corrections,
to the size of $\pi^\pm$.  For an electrically neutral, non-self-conjugate
meson $M_{i \bar j}$, the sign of $\langle r^2 \rangle$ is controlled by the
lighter of the $q_i$ and $\bar q_j$. For example, $\langle r^2 \rangle_{K^0} =
-(0.077 \pm 0.010)$ fm$^2$ \cite{pdg}, as can be understood from the larger
spatial extent of the $d$ compared with the $\bar s$, which results from the
property that $M_{ud} < M_s$.
\bibitem{r2ell}
%
More generally, $\langle r^{2\ell} \rangle = \int \rho(r) \, r^{2\ell} \,
d^3r$, and 
%
\beq
\langle r^{2\ell} \rangle = \frac{(2\ell+1)!}{\ell!} \, \frac{d^\ell F(q^2)}
{(dq^2)^\ell}{} \Big |_{q^2=0}
\label{r2ell}
\eeq
%
with our $(+,-,-,-)$ metric.
\bibitem{vform}
%
Combining Eqs. (\ref{v}) and (\ref{psi_scaling}), one has 
$V \propto r^\nu \propto \mu^{-\frac{\nu}{2+\nu}}$. 
Since the kinetic energy $E_{kin}$ and potential energy $E_{pot}$ satisfy
the virial relation $E_{kin}=(\nu/2)E_{pot}$, it follows that the total energy
$E$ satisfies 
$E \propto \left ( 1 + \frac{\nu}{2} \right )\mu^{-\frac{\nu}{2+\nu}}$
(negative for a bound state).
\end{thebibliography}
\end{document}